\documentstyle[twoside,fleqn,espcrc2,epsf]{article}
\title{
$K^{+}\to\pi^{+}\pi^{0}$ Decay Amplitude in Quenched Lattice 
QCD\thanks{presented by N. Ishizuka}
}
\author{
JLQCD Collaboration: 
S.~Aoki
\address{
Institute of Physics,
University of Tsukuba,
Tsukuba, Ibaraki 305, Japan
},
M.~Fukugita
\address{
Institute for Cosmic Ray Research,
University of Tokyo,
Tanashi, Tokyo 188, Japan
},
S.~Hashimoto
\address{
Computing Research Center,
High Energy Accelerator Research Organization (KEK),
Tsukuba, Ibaraki 305, Japan
},
N.~Ishizuka$^{\rm a,}$
\address{
Center for Computational Physics, 
University of Tsukuba,
Tsukuba, Ibaraki 305, Japan
},
Y.~Iwasaki$^{\rm a,d}$
K.~Kanaya$^{\rm a,d}$,
Y.~Kuramashi
\address{
Institute of Particle and Nuclear Studies,
High Energy Accelerator Research Organization (KEK),
Tsukuba, Ibaraki 305, Japan   
}
M.~Okawa$^{\rm e}$,
A.~Ukawa$^{\rm a}$,
T.~Yoshi\'{e}$^{\rm a,d}$
}
\begin{document}
%
%
\begin{abstract}
A new study is reported of a lattice QCD calculation of the 
$K^+\to\pi^+ \pi^0 $ decay amplitude with the Wilson quark action 
in the quenched approximation at $\beta=6.1$.
The amplitude is extracted from the $K\to\pi\pi$ Green function, and 
a conversion to the continuum value is made employing a recent one-loop 
calculation of chiral perturbation theory. 
The result is consistent with the experimental value if extrapolated 
to the chiral limit.  
\end{abstract}
\maketitle
%
%
\section{ Introduction }
It has long been a problem that the $\Delta I=3/2$ 
$K^+\to\pi^+\pi^0$ amplitude calculated in quenched lattice QCD is 
about a factor two too large compared to the experimental value\cite{BSReview}.
In this article we report a new study of this problem,
incorporating various theoretical and technical advances 
in recent years for analysis.
In particular we discuss in detail how one-loop corrections of 
chiral perturbation theory (CHPT) recently calculated\cite{ONECPTH} affect
physical predictions for the decay amplitude from lattice QCD simulations. 

Our simulation is carried out in quenched lattice QCD employing 
the standard plaquette action for gluons at $\beta=6.1$ and the Wilson action for quarks. 
Two lattice sizes, $24^3\times 64$ and $32^3\times 64$, are employed.
We take up, down and strange quarks to be degenerate,
and make measurements at four values of the common hopping parameter,
$\kappa = 0.1520$, $0.1530$, $0.1540$ and $0.1543$, which correspond to
$m_\pi/m_\rho = 0.797$, $0.734$, $0.586$ and $0.515$.
%
%
\section{ Extraction of decay amplitude }
The 4-quark operator most relevant for the 
$\Delta I=3/2$ $K\to\pi\pi$ decay is
$Q_+ = [ ( \bar{s} d )_L
         ( \bar{u} u )_L
    +    ( \bar{s} u )_L
         ( \bar{u} d )_L ]  / 2$.
We extract the decay amplitude from the 4-point correlation function defined by
$M_Q  = \langle 0 | W_+  W_0 Q_{+}(t) W_K | 0 \rangle$ 
where $W_{0,+,K}$ are wall sources for $\pi^0$, $\pi^+$ and $K^+$. 
In our calculations on a lattice of a temporal size $T=64$,   
the walls are placed at the time slices $t_{K^+}=4$, $t_{\pi^+}=59$ and $t_{\pi^0}=60$.
The mesons are all created at rest, and the 4-quark operator $Q_+$ is 
projected to zero spatial momentum.

We can use two types of factors to remove the normalization factors in $M_Q$.
If we define
\begin{eqnarray}
\lefteqn{ M_W  =
                   \langle 0 | W_0  \pi^{0} ( t ) | 0 \rangle
                   \langle 0 | W_+  \pi^{+} ( t ) | 0 \rangle
                   \langle 0 | K(t)   W_K         | 0 \rangle ,} && \nonumber \\
\lefteqn{ M_P  = 
    \langle 0 | K(t) W_K      | 0 \rangle 
    \langle 0 | W_+ W_0  \pi^{+} ( t ) \pi^{0} ( t ) | 0 \rangle , }  && 
\end{eqnarray}
we find
$R_W \equiv M_Q / M_W$
$\sim \langle \pi^+\pi^0 | Q_+ | K^+ \rangle$
$   / \langle \pi | \pi | 0 \rangle^3 \cdot {\rm exp}(t-t_+)\Delta$
and
$R_P \equiv M_Q / M_P$ 
$\sim \langle \pi^+\pi^0 | Q_+ | K^+ \rangle / \langle \pi | \pi | 0 \rangle^3$
for $t_K\ll t\ll t_+, t_0$, where $\Delta=m_{\pi\pi}-2m_\pi$ is the 
mass shift of the 2-pion state due to finite lattice size 
effects~\cite{LUSHER}.
 
\begin{figure}[t]
\vspace*{-0.2cm}
\centerline{\epsfxsize=7.0cm \epsfbox{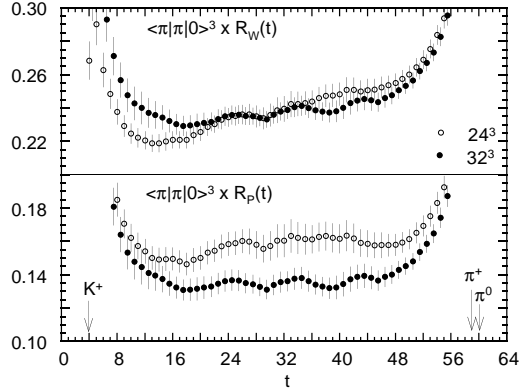}}
\vspace*{-1cm}
\caption{\label{RWFRPF}
$R_W \cdot \langle \pi | \pi | 0 \rangle^3$ and $R_P\langle \pi | \pi | 0 \rangle^3$
at $\kappa=0.153$.
Open and filled circles refer to data for $24^3$ and $32^3$ lattices.
}
\vspace*{-0.7cm}
\end{figure}

In Fig.~\ref{RWFRPF} we plot $\langle \pi | \pi | 0 \rangle^3 \cdot R_W$ and 
$\langle \pi | \pi | 0\rangle^3 \cdot R_P$ at 
$\kappa=0.1530$ as a function of time $t$ of the weak operator.
We clearly observe a non-vanishing slope for $R_W$, while $R_P$ exhibits a 
plateau as expected.
The decay amplitude can be obtained by fitting $R_W$ to a single exponential 
or $R_P$ to a constant. We find the results to be mutually consistent 
within the statistical error.
%
%
%
%
\begin{figure}[t]
\vspace{-0.6cm}
\centerline{\epsfxsize=7.0cm \epsfbox{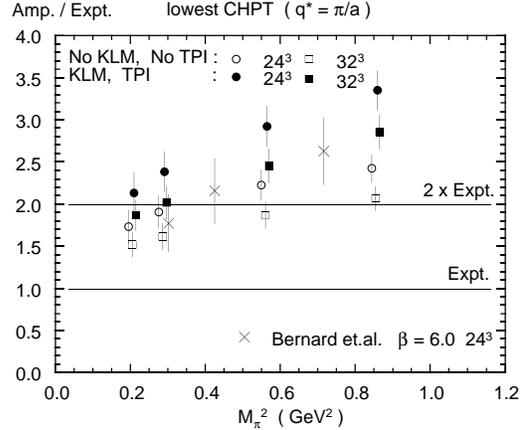}}
\vspace{-1.0cm}
\caption{ \label{COMPARISON}
Comparison of our results normalized by the experimental value obtained 
with tree-level CHPT relation for $q^*=\pi/a$ at $\beta=6.1$ 
with those of previous work{\protect\cite{BSKpipi}} at $\beta=6.0$ (crosses).
Open symbols refer to results with the traditional normalization 
$\protect\sqrt{2\kappa}$ and no tadpole improvement, 
and filled ones are those with the KLM normalization and tadpole improvement.
}
\vspace{-0.7cm}
\end{figure}
%
%

The one-loop renormalization factor relating the operator $Q_+$ on the lattice 
and that in the continuum was obtained in Refs.~\cite{Z-M,Z-BS}. 
We set $q^*=1/a$ or $\pi/a$ as the matching point and employ tadpole-improved
perturbation theory with 
the mean-field improved $\overline{\rm MS}$ coupling constant\cite{TPI}.  
Quark fields are normalized with the KLM factor~\cite{KLM}, 
$\sqrt{1-3\kappa/4\kappa_c}$.
%
%
\section{ Result }
In earlier calculations tree-level chiral perturbation theory (CHPT) 
formula was used to obtain the physical amplitude $A^{phys}$ from the 
lattice counterpart $A^{lat}$.  The formula takes the form 
$ A^{phys} = ( m_K^2 - m_\pi^2 )/( 2 M_\pi^2 ) \cdot A^{lat}$
where $m_K=497 {\rm MeV}$ and $m_\pi=136 {\rm MeV}$ are physical masses, 
and $M_\pi$ is the degenerate mass of $K$ and $\pi$ mesons on the lattice.
Clearly the physical amplitude should be independent of the 
lattice mass $M_\pi$ if the matching formula is exact.

In Fig.~\ref{COMPARISON} we compare results for the physical decay amplitude 
from a previous work\cite{BSKpipi} carried out at $\beta=6.0$ on a $24^3$ 
spatial lattice with those of our simulation at $\beta=6.1$. 
Our values obtained with the conventional quark normalization $\sqrt{2\kappa}$
(crosses and open squares) are consistent with their results,
but are larger than the experimental result roughly by a factor two.

Let us also note with our results (circles and squares) 
that (i) the KLM normalization and tadpole improvement of 
the operator have a significant effect on the amplitude, (ii) there is 
a clear dependence of the amplitude on the lattice meson mass $M_\pi$, 
and (iii) a significant finite-size effect is observed between the two 
lattice sizes.
These features show that the tree-level CHPT is inadequate to extract 
the physical amplitude from lattice results.

\begin{figure}
\vspace*{-0.6cm}
\centerline{\epsfxsize=7.0cm \epsfbox{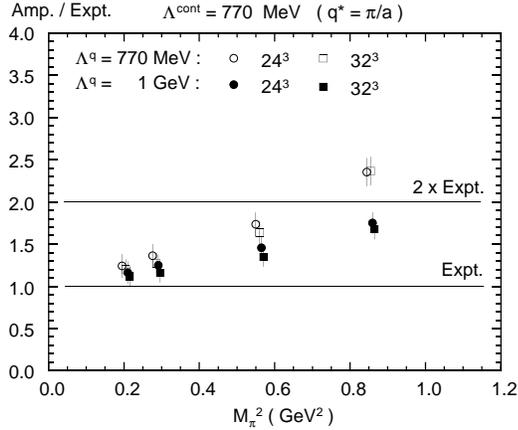}}
\vspace*{-1cm}
\caption{ \label{FRHO} 
Decay amplitude normalized by experimental value for $q^*=\pi/a$ 
obtained with one-loop CHPT for $\Lambda^{cont} = 770{\rm MeV}$.
Circles and squares  refer to data for $24^3$ and $32^3$ spatial sizes. 
Open symbols are for $\Lambda^{q}=770{\rm MeV}$ and filled ones for 
$1\ {\rm GeV}$.
}
\vspace*{-7mm}
\end{figure}

In Fig.~\ref{FRHO} we show how predictions for the physical amplitude 
change if we apply the one-loop formula of CHPT~\cite{ONECPTH} to our results.
Here $\Lambda^q$ and $\Lambda^{cont}$ are the cutoffs of CHPT for quenched and full theory.
In converting to physical values we ignore effects of $O(p^4)$ contact terms 
of the CHPT Lagrangian since their values are not well known.
An interesting point is that a size dependence seen with 
the tree-level analysis in Fig.~\ref{COMPARISON} is absent.
Another important point is that the magnitude of the amplitude decreases 
by $30-40$\% over the range of meson mass covered in our simulation.
The amplitude depends significantly on the quenched lattice cutoff 
$\Lambda^q$, in particular toward larger values of $M_\pi$.

%
%
%
%
Our results in Fig.~\ref{FRHO} show that a sizable meson mass dependence still remains.
This may be attributed to $O(p^4)$ contact terms which were not taken into account.
If we assume that the contact terms of the full theory are as small 
as suggested by the phenomenological analysis~\cite{BK},
we can estimate the physical amplitude by a linear extrapolation of $M_\pi^2$ to the chiral limit.
In Table~\ref{FINALFIT} we list results for several choices of the cutoff
and the operator matching point $q^*$.
As is expected from Fig.~\ref{FRHO}, variation of results on the 
choice of the cutoff parameters is small, being of the order of 10\%.
Allowing for this uncertainty, the values in Table~\ref{FINALFIT}
are consistent with the experimental value of $10.4\times 10^{-3} {\rm GeV}^3$.
%
%
\section{ Conclusions }
The present study has shown that the quenched result for the 
$K^+\to\pi^+ \pi^0 $ decay amplitude agrees with experiment much better 
than previously thought, especially if the chiral limit is taken 
for the lattice meson mass.
Away from the chiral limit, the amplitude depends significantly on the cutoff scales of CHPT, however.  
This problem is largely ascribed to a mismatch of chiral logarithms 
between the quenched and full theories\cite{ONECPTH}.  
In order to obtain $K^+\to\pi^+ \pi^0 $ decay amplitude free from this uncertainty,
we need simulations in full QCD.

%
%
\begin{table}[t]
\setlength{\tabcolsep}{0.2pc}
\begin{tabular}{llllll}
\hline
\hline
&&\multicolumn{4}{c}{$C_+\langle \pi\pi |Q_+| K \rangle (\times 10^{-3} {\rm GeV}^3)$}\\
$\Lambda^{cont}$   &  $\Lambda^{q}$   &  \multicolumn{2}{c}{$24^3$}&\multicolumn{2}{c}{$32^3 $}\\
({\small GeV}) &  ({\small GeV}) &$q^*=\frac{1}{a}$&$q^*=\frac{\pi}{a}$&$q^*=\frac{1}{a}$&$q^*=\frac{\pi}{a}$\\[1mm]
\hline
0.77  &  0.77  &  $ 9.3(19)$&$10.2(21)$  & $8.9(17)$&$ 9.7(19)$ \\
0.77  &  1.0   &  $ 9.4(13)$&$10.3(14)$  & $8.8(11)$&$9.6(12)$ \\
1.0   &  0.77  &  $10.3(21)$&$11.3(23)$  & $9.8(19)$&$10.7(21)$ \\
1.0   &  1.0   &  $10.4(14)$&$11.4(15)$  & $9.7(12)$&$10.6(13)$ \\
\hline
\hline
\end{tabular}
\vspace*{3mm}
\caption{\label{FINALFIT}
Results of linear extrapolation of the decay amplitude to $M_\pi^2=0$. 
Statistical and extrapolation errors are combined.
The experimental value is $10.4\times 10^{-3} {\rm GeV}^3$.
}
\vspace*{-7mm}
\end{table}
%
%
\hfill\break
This work is supported by the Supercomputer 
Project (No.~97-15) of High Energy Accelerator Research Organization (KEK),
and also in part by the Grants-in-Aid of 
the Ministry of Education (Nos. 08640349, 08640350, 08640404,
09246206, 09304029, 09740226).
%
%

%

\begin{thebibliography}{9}
%
%
\bibitem{BSReview} 
C. Bernard and A. Soni, Nucl. Phys. {\bf B}(Proc. Suppl.){\bf 9} (1989) 155.

\bibitem{BSKpipi} 
C. Bernerd and A. Soni, Nucl. Phys. {\bf B}(Proc. Suppl.){\bf 17} (1990) 495.
%
\bibitem{ONECPTH} M.F.L. Golterman and K. C. Leung, hep-lat/9702015.
%
\bibitem{LUSHER} M. L\"uscher, Comm. Math. Phys. {\bf 105} (1986) 153.
%
\bibitem{Z-M} G. Martinelli, Phys. Lett. {\bf 141B} (1984) 395.
%
\bibitem{Z-BS} C. Bernard, T. Draper, and A. Soni, Phys. Rev. {\bf D36} (1987) 3224.
%
\bibitem{TPI} G.P. Lepage and P.B. Mackenzie, Phys. Rev. {\bf D48} (1993) 2250.
%
%
\bibitem{KLM} G.P. Lepage, Nucl. Phys. {\bf B26} (Proc.Suppl.) (1992) 45; 
 P. Mackenzie, Nucl. Phys. {\bf B34} (Proc.Suppl.) (1994) 35;
 A. Kronfeld, Nucl. Phys. {\bf B34} (Proc.Suppl.) (1994) 415.
%
%
%
%
%
%
%
%
%
%
%
%
\bibitem{BK}
J. Bijnens, H. Sonoda and M. B. Wise, Phys. Rev. Lett. {\bf 53} (1984) 2367;
J. Kambor, J. Missimer and D. Wyler, Phys. Lett. {\bf B261} (1991) 496.
%
\end{thebibliography}
\end{document}